\documentclass{PoS}

\usepackage{xspace}
\usepackage{cite}

\newcommand{\wz}{\texttt{WHIZARD}\xspace}
\newcommand{\om}{\texttt{O'Mega}\xspace}
\newcommand{\vamp}{\texttt{VAMP}\xspace}
\newcommand{\circe}{\texttt{CIRCE}\xspace}
\newcommand{\go}{\texttt{GoSam}\xspace}
\newcommand{\ol}{\texttt{OpenLoops}\xspace}
\newcommand{\si}{\texttt{SINDARIN}\xspace}
\newcommand{\mad}{\texttt{MG5\_amC@NLO}\xspace}
\newcommand{\olp}{\texttt{.olp}\xspace}
\newcommand{\olc}{\texttt{.olc}\xspace}
\title{Automated NLO QCD Corrections with \wz}

\ShortTitle{Automated NLO QCD Corrections with \wz}

\author{\speaker{Christian Weiss}\\%\thanks{A footnote may follow.}\\
        DESY Theory Group, Notkestr. 85, D-22607 Hamburg, Germany\\
        University of Siegen, Department of Physics, Walter-Flex-Str. 3, D-57068 Siegen, Germany\\
        E-mail: \email{christian.weiss@desy.de}}

\author{Bijan Chokouf\'{e} Nejad\\
        DESY Theory Group, Notkestr. 85, D-22607 Hamburg, Germany\\
        E-mail: \email{bijan.chokoufe@desy.de}}

\author{Wolfgang Kilian\\
	University of Siegen, Department of Physics, Walter-Flex-Str. 3, D-57068 Siegen, Germany\\
	E-mail: \email{kilian@physik.uni-siegen.de}}
	
\author{J\"urgen Reuter\\
	DESY Theory Group, Notkestr. 85, D-22607 Hamburg, Germany\\
	E-mail: \email{juergen.reuter@desy.de}}

\abstract{We briefly discuss the current status of NLO QCD automation in the Monte Carlo event generator \wz. The functionality is presented for the explicit study of off-shell top quark production with associated backgrounds at a lepton collider.\\
 \begin{flushright}
    \normalsize{} DESY 15--175
 \end{flushright}}

\FullConference{The European Physical Society Conference on High Energy Physics\\
		22--29 July 2015\\
		Vienna, Austria}

\begin{document}

\section{The \wz Event Generator}
\wz \cite{Kilian:2007gr} is a multi-purpose event generator for in principle arbitrary processes at hadron and lepton colliders. It is especially suited for lepton collider physics due to its generic treatment of beam-spectra and initial-state photon radiation. Moreover, the simulation and analysis can be very conveniently performed using the built-in script language \si.\\
\wz is written in Fortran2003. Its structure is strictly object-oriented, so that a modular structure enables
the convenient interface to numerous other programs. The main sub-components of \wz are \om\cite{Moretti:2001zz}, \vamp\cite{Ohl:1998jn} and \circe\cite{Ohl:1996fi}:\\
\om provides multi-leg tree-level matrix elements using the helicity formalism. \vamp is used for Monte-Carlo integration and grid sampling. It is 
a multi-channel version of usual adaptive integration methods. The \texttt{CIRCE} packages create and evaluate lepton beam spectra.\\
Scattering amplitudes and color factors are factorized in \wz. Color factors are evaluated using the color-flow formalism \cite{Kilian:2012pz}.\\
\wz can be used for event generation on parton level as well as for the subsequent shower simulation. For this purpose, it has its own analytical parton shower \cite{Kilian:2011ka} as well as a built-in interface to \texttt{Pythia6} \cite{Sjostrand:2006za}. An automated interface to \texttt{Pythia8} \cite{Sjostrand:2007gs} or \texttt{HERWIG++} \cite{Bahr:2008pv} is not yet present, but planned.\\
The list of external packages that can be linked to \wz covers the most commonly used high-energy physics simulation
and analysis tools, such as \texttt{FastJet} \cite{Cacciari:2011ma}, \texttt{LHAPDF} \cite{Buckley:2014ana} or \texttt{HepMC} \cite{Dobbs:2001ck}. For next-to-leading order studies, \go \cite{Cullen:2014yla} or \ol \cite {Cascioli:2011va} can be used to compute virtual loop matrix elements.\\
Though \wz spearheaded many BSM phenomenological studies \cite{Kilian:2004pp, Hagiwara:2005wg, Beyer:2006hx, Alboteanu:2008my, Kalinowski:2008fk, Kilian:2014zja} we will focus here on automation of SM QCD NLO corrections.
%%%%%%%%%%%%%%%%%%%%%%%%%%%%%%%%%%%%%%%%%%%%%%%%%%%%%%%%%%%%%%%
\section{Automated NLO Calculation with \wz}
Next-to-leading order (NLO) calculations have become standard for the prediction of most observables during the last decade. Substantial progress has been made in the automation of loop matrix-element computation as well as in NLO parton shower matching \cite{Nason:2004rx, Frixione:2002ik}. At the LHC, NLO simulations are routinely employed.\\
Computer programs in this field cover a range from dedicated, single-purpose codes to automated multi-purpose event generators. In the latter category, NLO support has so far been achieved by \texttt{Madgraph} \cite{Alwall:2014hca} and \texttt{Sherpa} \cite{Hoche:2010pf}.\\
With the ILC approaching its possible approval phase, ILC studies become more specific \cite{Baer:2013cma}, thereby increasing the need for easy to use NLO programs increases. Here, we want to combine the expertise of \wz in the field of lepton collisions with the improved accuracy of NLO predictions.\\
There have been earlier works on NLO QED extensions to \wz for certain supersymmetric processes \cite{Kilian:2006cj, Robens:2008sa} as well as on NLO QCD corrections for $pp \rightarrow b\bar{b}b\bar{b}$ \cite{Binoth:2009rv, Greiner:2011mp} using Catani-Seymour subtraction. However, the fully generic NLO framework is a recent devolopment.
\subsection{Subtraction schemes}
The main task about an automated NLO framework is to treat the UV- and IR-divergences which come along with loop matrix elements and real-emission amplitudes. Whereas UV-divergences can be absorbed via renormalization into the physical quantities of the underlying field theory, IR-divergences only cancel in the sum of real and virtual matrix elements, as stated by the KLN theorem \cite{Kinoshita:1962ur, Lee:1964is}. However, this requires the choice of a regularization scheme, e.g. a lower cutoff parameter like gluon masses, or, most commonly, dimensional regularization.\\
The latter method makes divergences explicit by extending the number of integration dimensions to an arbitrary (complex) number different from the four physical ones. \\
Subtraction schemes are a different approach especially suited for numerical calculations. Additional subtraction terms $\cal{C}$, which reproduce the singularities of the real and virtual matrix elements, are added to the NLO cross section,
such that
\begin{equation}
	d\sigma^{\rm{NLO}} = d\sigma^{\rm{LO}} + \underbrace{\int_{n+1}\left(d\sigma^{\rm{R}} - d\sigma^{\rm{S}}\right)}_{\rm{finite}} 
                      + \underbrace{\int_{n+1} d\sigma^{\rm{S}} + \int_n d\sigma^{\rm{V}}}_{\rm{finite}}.
   \label{subtraction_def}
\end{equation}
The explicit form of $\mathcal{C}$ is arbitrary. Several approaches have been developed, the most popular ones
being the Catani-Seymour scheme \cite{Catani:1996vz} and the FKS (Frixione-Kunszt-Signer) scheme \cite{Frixione:1995ms, Frixione:2009yq}, described below. Catani-Seymour subtraction has been the standard method for the last two decades, due to the fact that it can be used with any phase space generator. Thus, it is especially suited for process-specific event generators and is also widely used in this context. However, the FKS scheme has increased in popularity over last ten years and is also used by \wz. Other programs with their own FKS-implementation are \mad, \texttt{HERWIG++}, \texttt{HELAC-NLO} \cite{Bevilacqua:2011xh} and the \texttt{POWHEG-BOX} \cite{Alioli:2010xd}. All of them are automated or semi-automated event generators, which allow for the coherent use of a single phase space generator, suited for FKS. Also, the number of CS dipoles grows larger than the number of FKS regions. Moreover, it is known that the Catani-Seymour procedure requires an increased complexity when NLO parton shower matching, e.g. with the \texttt{POWHEG} method, is performed \cite{Muck:2015cra}.\\
%%%%%%%%%%%%%%%%%%%%%%%%%%%%%%%%%%%%%%%%%%%%%%%%%
\subsection{BLHA interface}
\wz with \om matrix elements can only generate tree-level processes. External programs can be interfaced to \wz, so-called One-Loop Providers (OLPs), to obtain virtual matrix elements. To standardize the interface between OLP and Monte-Carlo program, the Binoth Les Houches Accord (BLHA) \cite{Binoth:2010xt, Alioli:2013nda} has been developed. It specifies standards for the communication files between OLP and the Monte-Carlo program and also explicitly prescribes the functions with which the OLP program can be called, mostly to compute matrix elements.\\
\wz has a generic BLHA-interface. This means that OLPs which comply with the BLHA standards can be linked to the main program with only minor programming effort. For a working interface, \wz will produce an \olp-file, which contains general information about the process, e.g. the type of correction (QCD or electroweak) or the desired regularisation scheme, as well as the involved flavor structures and the corresponding amplitude types (tree, color-correlated tree, spin-correlated tree and loop). The One-Loop Provider then reads in the \olp-file and, if successful, will provide the necessary libraries, which can be linked to \wz. It also generates a contract(\olc)-file, which in turn is read by \wz. The \olc-file contains information about the interfaces to the BLHA-library.\\
Up to now, \wz has working BLHA interfaces for two programs, \go and \ol.\\
\go is an OLP creating explicit analytical code based on Feynman diagrams using \texttt{QGraf} \cite{Nogueira:1991ex} and \texttt{FORM} \cite{Kuipers:2012rf}. This makes \go a very flexible program, suited for SM as well as for BSM processes. Virtual diagrams are evaluated using $D$-dimensional reduction \cite{Ossola:2006us} or tensorial reconstruction \cite{Mastrolia:2008jb}.\\
On the other hand, \ol uses recursion relations to compute matrix elements. The desired process libraries have to be installed from the \ol-repository. This introduces the drawback that only these pre-compiled libraries can be used. However, this is made up for by the increased speed of the program, which is mainly due to the faster matrix-element evaluation, but also because the user does not have to compile the libraries by himself as is the case with \go.
%%%%%%%%%%%%%%%%%%%%%%%%%%%%%%%%%%%%%%%%%%%%%%%%%%%%%%%%%%%%%%%%%%%
\section{Top Quark off-shell Production and Associated Background}
\begin{figure}
	\begin{center}
	   \includegraphics[scale=.5]{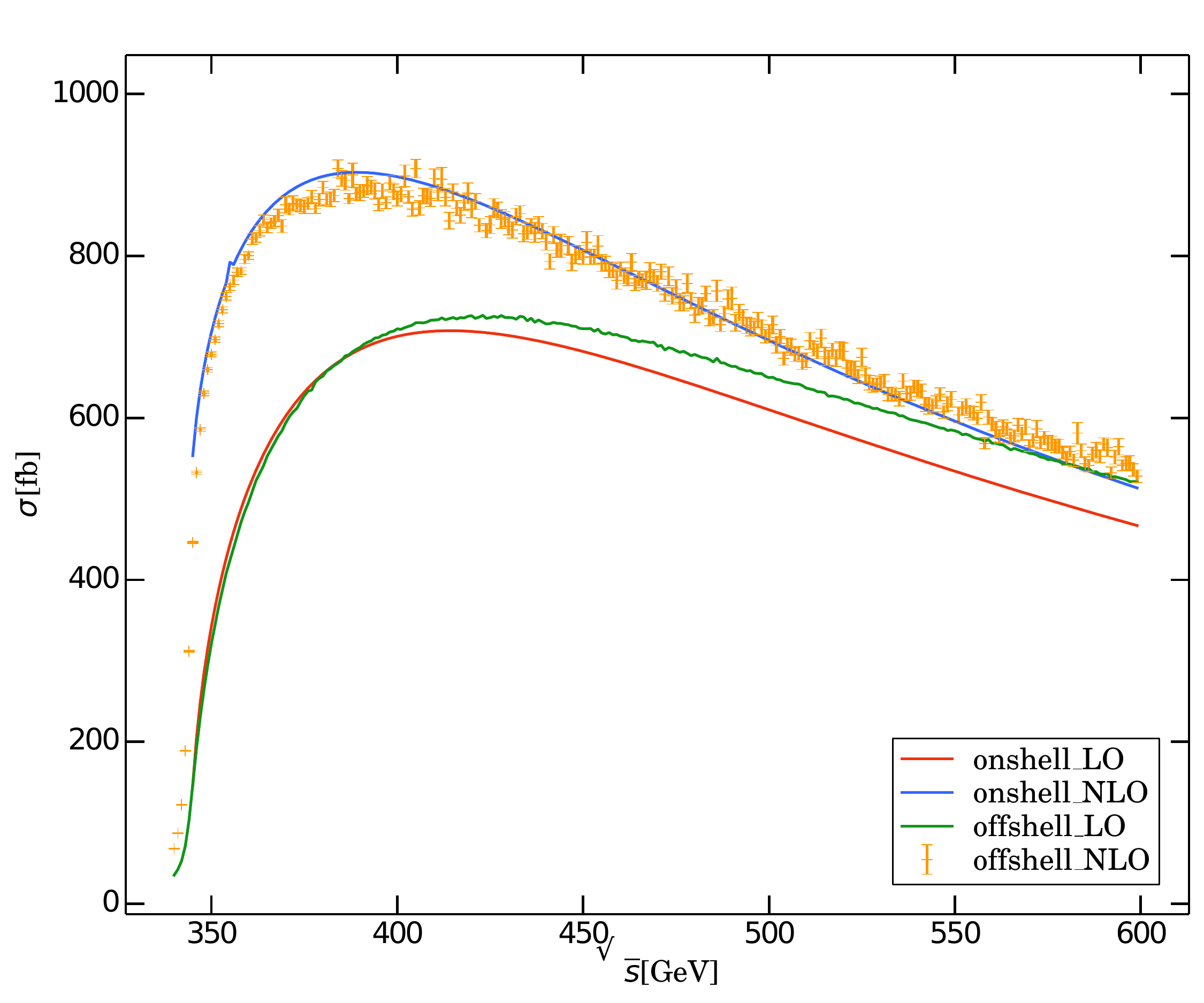}
	   \caption{The total cross section for the processes $e^+ e^- \rightarrow t \bar{t}$ and $e^+ e^- \rightarrow W^+ W^- b \bar{b}$ at LO and NLO.}
	   \label{TotalXsection}
	\end{center}
\end{figure}

In this section, we present a phenomenological study of the process $e^+ e^- \rightarrow W^+ W^- b\bar{b}$ at next-to-leading order QCD for a future lepton collider. A comprehensive study of this process for hadron colliders is given in \cite{Denner:2012yc}.\\
This signature is dominated by top-quark pair production and the subsequent decay $t \rightarrow bW$, but also contains the
process $e^+ e^- \rightarrow HZ$, which yields the same final state for the subsequent decays $H \rightarrow b\bar{b}$ and $Z \rightarrow W^+ W^-$. Both processes have never been measured at a lepton collider. However, future lepton colliders will definitely have the necessary center-of-mass energy and luminosity. Especially the ILC is designed to also operate at the top-producton threshold. Thus, a detailed study of top-quark properties will be possible.\\
A first numerical approach to this process was made by Ref. \cite{Lei:2008ii}, focussing on Higgs mass effects using a cut-off regularization specialized on this process. However, we were not able to reproduce both their LO- and NLO-results with \wz as well as with \mad.\\
We present the preliminary results for the total cross section at next-to-leading order as well as fixed-order NLO event shapes for selected observables. 
Our calculation was performed with the release version 2.2.7 of \wz, using the standard NLO-setup. The loop matrix elements are obtained from \texttt{OpenLoops}. \\
For our setup, we use $m_b = 4.7\, \rm{GeV}$ and $m_t = 172\, \rm{GeV}$. The renormalization scale is chosen as $\mu_R = m_t$. We set $\alpha_e^{-1} = 132.160$ and $\alpha_s(M_Z) = 0.118$. All amplitudes are computed using the top-quark width at NLO for massive b-quarks, as computed in \cite{Jezabek:1988iv}. The LO cross section, accordingly, is obtained using the well-known LO value for $\Gamma_t$. This yields $\Gamma_t^{\rm{NLO}} = 1.409\, \rm{GeV}$ and $\Gamma_t^{\rm{LO}} = 1.538\, \rm{GeV}$.\\ 
Figure \ref{TotalXsection} shows the total cross section for the process $e^+ e^- \rightarrow W^+ W^- b\bar{b}$ at leading and next-to-leading order. For comparison, also the result for on-shell $t\bar{t}$-production is displayed. It can be seen that the simple assumption $\sigma_{\rm{offshell}} / \sigma_{\rm{onshell}} \approx BR (t \rightarrow bW) \approx 1$ can not be applied everywhere due to background contributions. %for small $\sqrt{s}$, the ratio of the on-shell and off-shell LO cross section corresponds to the branching ratio $\Gamma(t \rightarrow bW)$. However, the curves deviate for increasing values of $\sqrt{s}$ due to increasing background contributions.\\
We see that the NLO curves display the same behaviour. However, the intersection of both curves is at higher values of $\sqrt{s}$, and the difference between on-shell and off-shell results is not as distinct after this intersection as it is in the LO-case.
The K-factor close to the $t\bar{t}$-threshold can go up to almost 3. This large value is due to non-relativistic top quarks that can form a quasi-bound state via the exchange of Coulomb gluons. This leads to a contribution of large logarithms which have to be treated with resummation \cite{Hoang:2011it, Bach:2014hca}.\\
\begin{figure}
	\begin{center}
		\includegraphics[scale=.55]{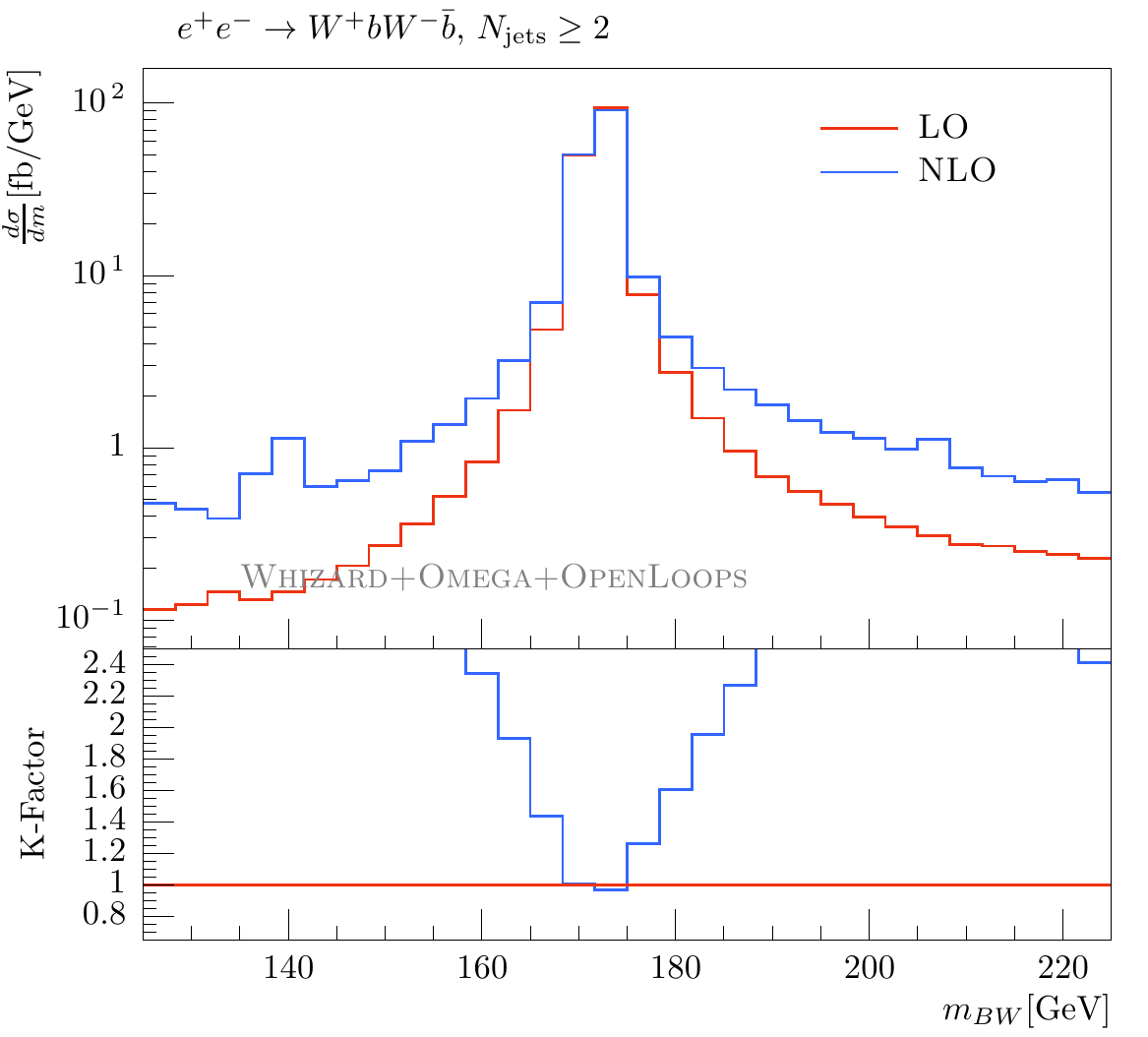}
		\hspace{1cm}
		\includegraphics[scale=.55]{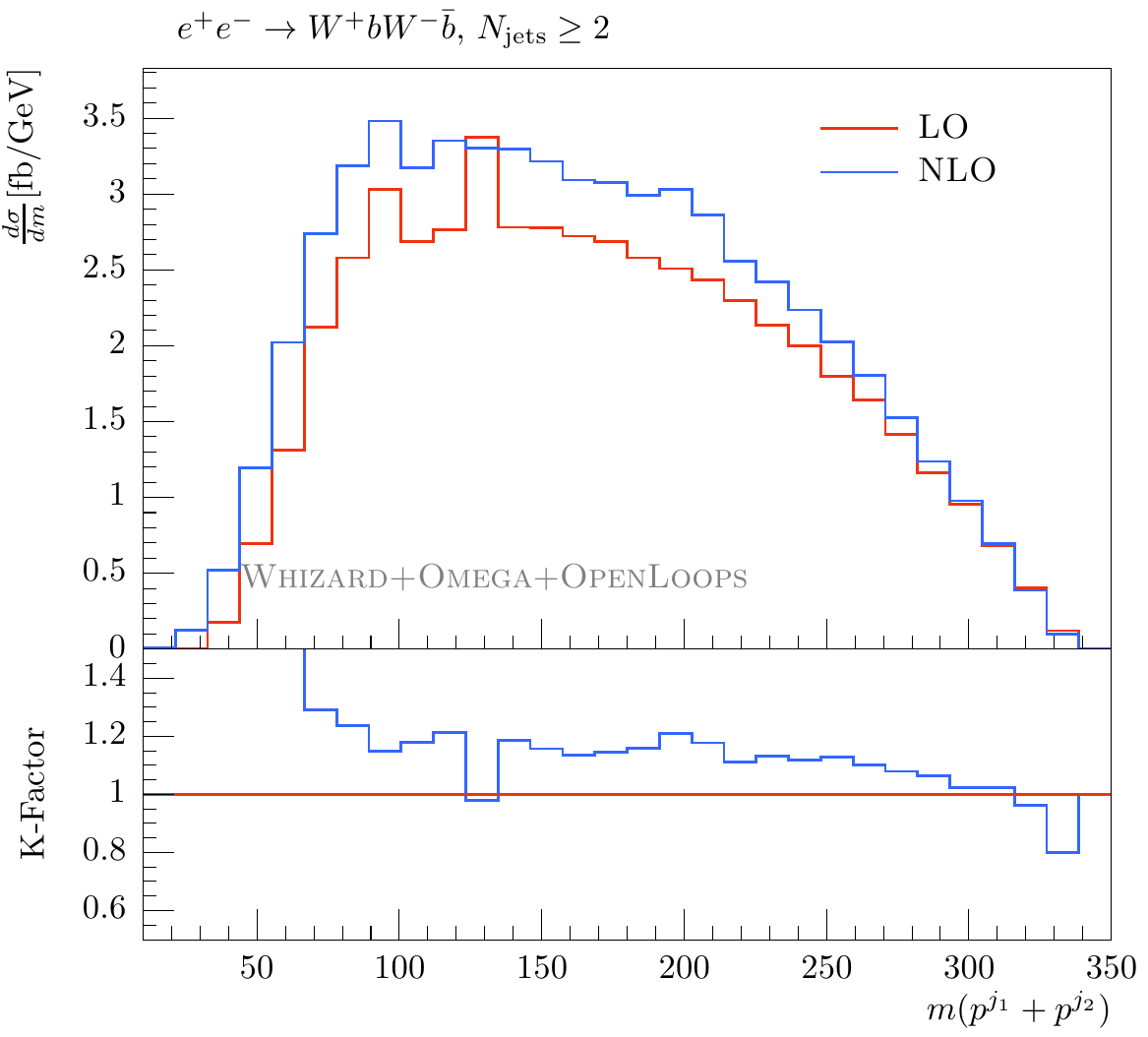}\\
		\includegraphics[scale=.55]{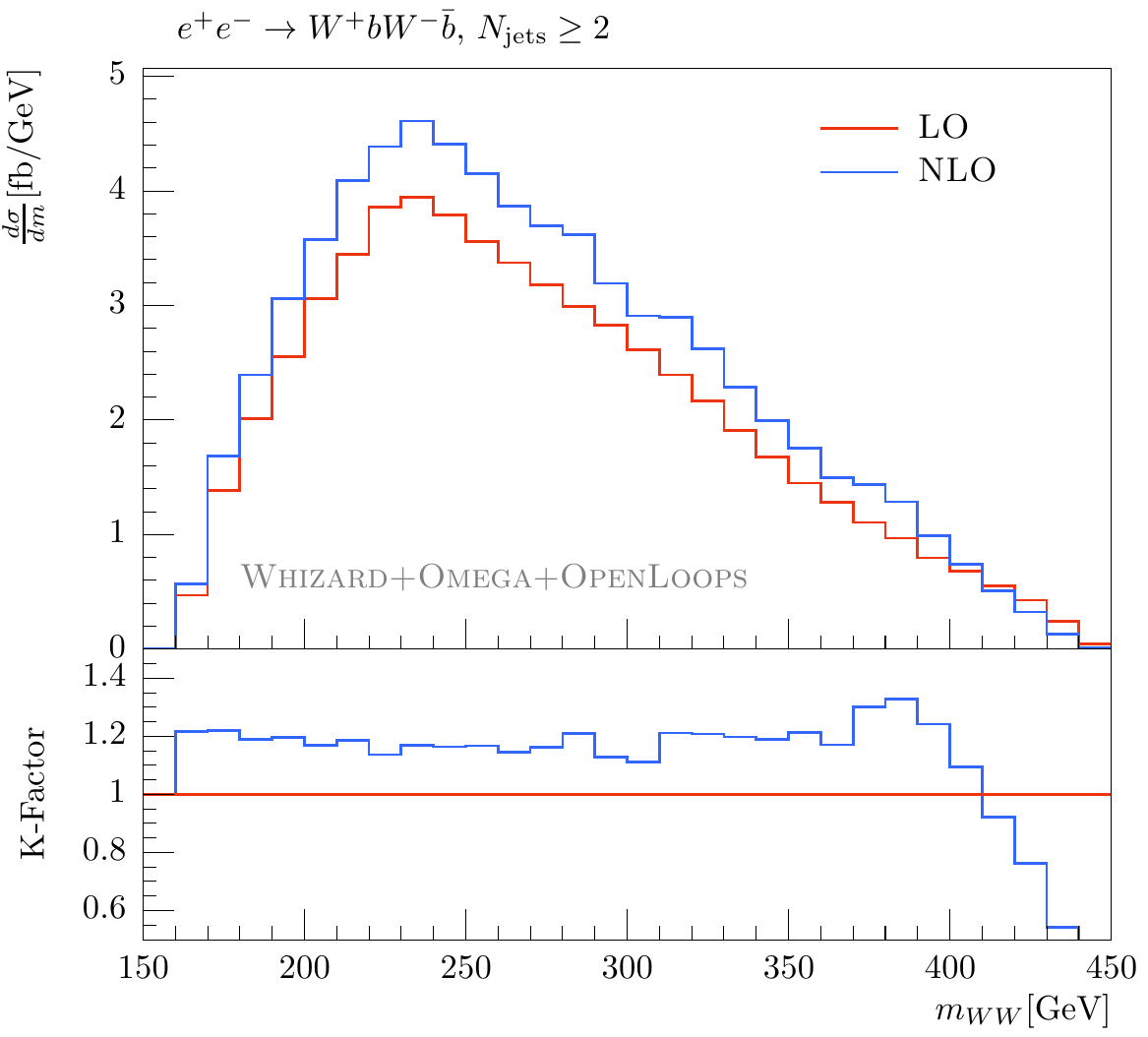}
		\hspace{1cm}
		\includegraphics[scale=.55]{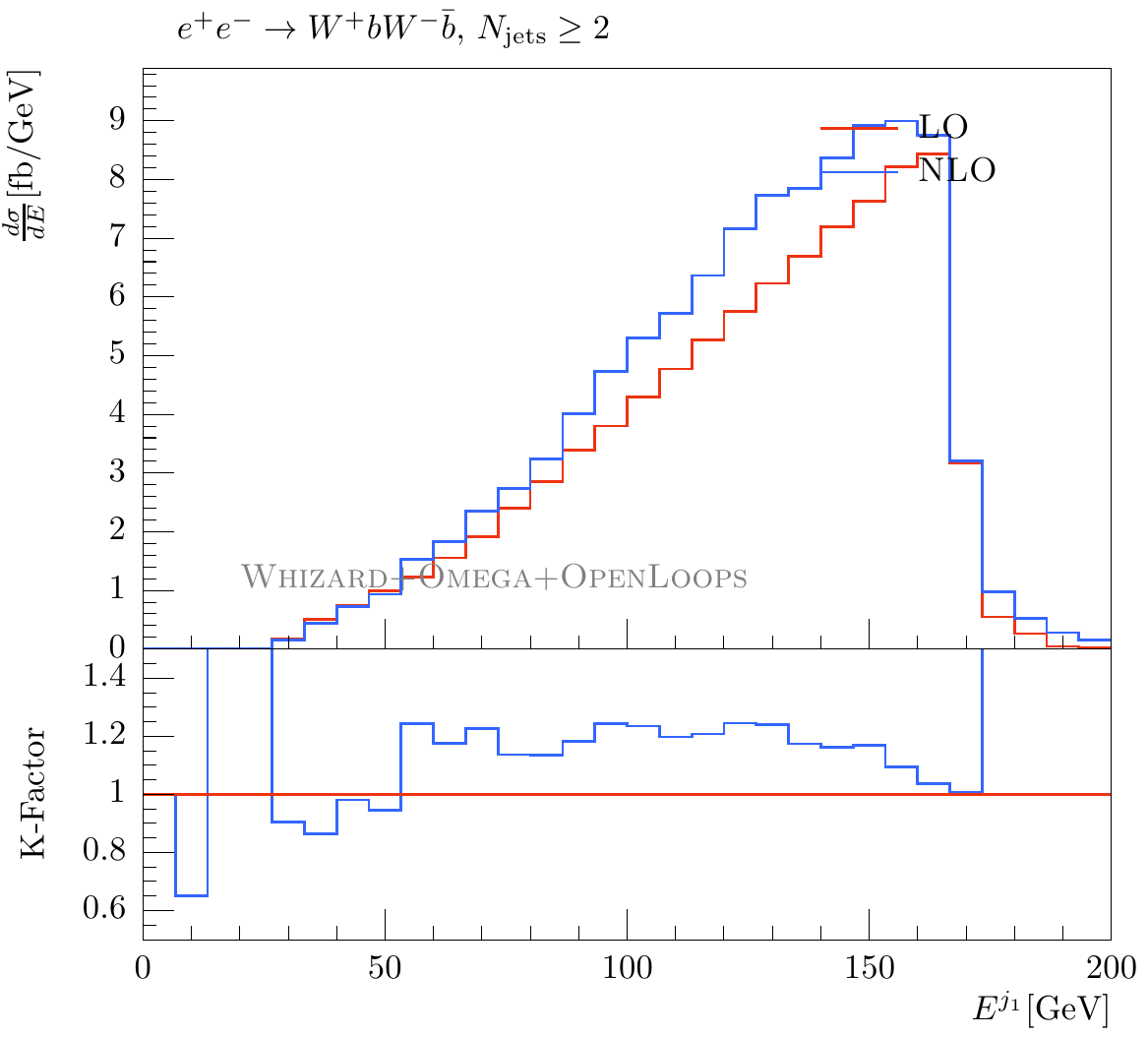}
		\caption{Various observables at $\sqrt{s} = 500\, \rm{GeV}$.
		   Top left: Invariant mass of $bW$-pairs;
	           Top right: Invariant mass of the first- and second-hardest jets;
	           Bottom left: Invariant mass of $WW$-pairs;
	           Bottom right: Leading-jet energy.}
		\label{events}
	\end{center}
\end{figure}
Figure \ref{events} displays several observables obtained from a simulated LO and NLO event sample. The LO sample contains 500K unweighted events. The NLO event sample consists of 45M weighted events. Of these, 15M have Born kinematics and their associated weight is $\mathcal{B} + \mathcal{V} - \sum_\alpha \mathcal{C_\alpha}$. Each singular region $\alpha$ gets associated with a real-emission event with weight $\mathcal{R}_\alpha$, thus giving the additional 30M events in the NLO sample.\\
The event samples are stored in the \texttt{HepMC}-format. We have used a variant of the anti-$k_t$ algorithm \cite{Cacciari:2008gp} with $\Delta R = 1$ to cluster jets, in this case b-quarks and gluons with $E > 1\, \rm{GeV}$, and required at least two jets to be present after clustering. The analysis was performed with \texttt{Rivet} \cite{Buckley:2010ar}.\\
In the $t\bar{t}$-production channel, the $bW$-pairs originate from the decay of the individual top quarks, so there is a clear resonance at $m_t = 172\, \rm{GeV}$. We see that this resonance is not affected significantly by NLO corrections. However, outside the resonance region, background contributions increase. We observe a similar enhancement in the other observables.\\
It must be noted that the events displayed in Figure \ref{events} have not been processed further with a parton shower algorithm. However, this will lead to unphysical results because double-counting occurs. To circumvent this, \wz has its own implementation of the Powheg method, which is presented in \cite{Chokoufe:2015xx}.
\section{Summary and Outlook}
We have shown that \wz is capable of performing full NLO QCD calculations in an automated way. For this purpose, \wz uses the FKS subtraction scheme. External BLHA one-loop providers can be used, primarily for the computation of loop amplitudes, but also for the evaluation of tree-level diagrams. The latter is useful to filter leading-order diagrams if \go is used. \\
We have presented phenomenological results for the specific example of off-shell top quark pair production with its associated backgrounds. We observe that NLO corrections can become large in certain distributions even if the correction to the total cross section is small.\\
Next-to-leading order simulations in \wz are still an experimental feature and not fully validated. Additionally, the focus of the development is on NLO QCD corrections for lepton collsions. Hadron collisions involve additional initial-state corrections and corrections to parton density functions. This feature will be supported once the support for lepton collisions in \wz is complete.\\
\wz will expand its expertise for lepton colliders at NLO via the consistent inclusion of beam structure functions at next-to-leading order. Moreover, since ILC beams will be polarized, there will be the possibility to include virtual matrix elements for explicit polarisations in the near future (up to now, BLHA loop providers deliver helicity-averaged matrix elements).\\
\acknowledgments
We are grateful to N. Greiner for many helpful discussions on NLO event generation in general as well as technical support with \go. Also, we thank J. Lindert for providing a variety of before non-available \ol-features, especially the required loop libraries for the processes studied in this paper. We also appreciate the help of both of them about interfacing their respective programs to \wz.


\begin{thebibliography}{99}
\bibliographystyle{unsrt}
\bibitem{Kilian:2007gr}
 W.~Kilian, T.~Ohl and J.~Reuter,
 %``WHIZARD: Simulating Multi-Particle Processes at LHC and ILC,''
 Eur.\ Phys.\ J.\ C {\bf 71}, 1742 (2011)
 [arXiv:0708.4233 [hep-ph]].
 %%CITATION = ARXIV:0708.4233;%%
 %339 citations counted in INSPIRE as of 08 Oct 2015
\bibitem{Moretti:2001zz}
  M.~Moretti, T.~Ohl and J.~Reuter,
  %``O'Mega: An Optimizing matrix element generator,''
  In *2nd ECFA/DESY Study 1998-2001* 1981-2009
  [hep-ph/0102195].
  %%CITATION = HEP-PH/0102195;%%
  %244 citations counted in INSPIRE as of 08 Oct 2015
\bibitem{Ohl:1998jn}
  T.~Ohl,
  %``Vegas revisited: Adaptive Monte Carlo integration beyond factorization,''
  Comput.\ Phys.\ Commun.\  {\bf 120}, 13 (1999)
  [hep-ph/9806432].
  %%CITATION = HEP-PH/9806432;%%
  %58 citations counted in INSPIRE as of 08 Oct 2015
\bibitem{Ohl:1996fi}
  T.~Ohl,
  %``CIRCE version 1.0: Beam spectra for simulating linear collider physics,''
  Comput.\ Phys.\ Commun.\  {\bf 101}, 269 (1997)
  [hep-ph/9607454].
  %%CITATION = HEP-PH/9607454;%%
  %112 citations counted in INSPIRE as of 08 Oct 2015
\bibitem{Kilian:2012pz}
  W.~Kilian, T.~Ohl, J.~Reuter and C.~Speckner,
  %``QCD in the Color-Flow Representation,''
  JHEP {\bf 1210}, 022 (2012)
  [arXiv:1206.3700 [hep-ph]].
  %%CITATION = ARXIV:1206.3700;%%
  %12 citations counted in INSPIRE as of 08 Oct 2015
\bibitem{Kilian:2011ka}
  W.~Kilian, J.~Reuter, S.~Schmidt and D.~Wiesler,
  %``An Analytic Initial-State Parton Shower,''
  JHEP {\bf 1204}, 013 (2012)
  [arXiv:1112.1039 [hep-ph]].
  %%CITATION = ARXIV:1112.1039;%%
  %18 citations counted in INSPIRE as of 08 Oct 2015
\bibitem{Sjostrand:2006za}
  T.~Sjostrand, S.~Mrenna and P.~Z.~Skands,
  %``PYTHIA 6.4 Physics and Manual,''
  JHEP {\bf 0605}, 026 (2006)
  [hep-ph/0603175].
  %%CITATION = HEP-PH/0603175;%%
  %6768 citations counted in INSPIRE as of 08 Oct 2015
\bibitem{Sjostrand:2007gs}
  T.~Sjostrand, S.~Mrenna and P.~Z.~Skands,
  %``A Brief Introduction to PYTHIA 8.1,''
  Comput.\ Phys.\ Commun.\  {\bf 178}, 852 (2008)
  [arXiv:0710.3820 [hep-ph]].
  %%CITATION = ARXIV:0710.3820;%%
  %1898 citations counted in INSPIRE as of 08 Oct 2015
\bibitem{Bahr:2008pv}
  M.~Bahr {\it et al.},
  %``Herwig++ Physics and Manual,''
  Eur.\ Phys.\ J.\ C {\bf 58}, 639 (2008)
  [arXiv:0803.0883 [hep-ph]].
  %%CITATION = ARXIV:0803.0883;%%
  %1047 citations counted in INSPIRE as of 08 Oct 2015
\bibitem{Cacciari:2011ma}  
  M.~Cacciari, G.~P.~Salam and G.~Soyez,
  %``FastJet User Manual,''
  Eur.\ Phys.\ J.\ C {\bf 72}, 1896 (2012)
  [arXiv:1111.6097 [hep-ph]].
  %%CITATION = ARXIV:1111.6097;%%
  %1010 citations counted in INSPIRE as of 08 Oct 2015
\bibitem{Buckley:2014ana}
  A.~Buckley, J.~Ferrando, S.~Lloyd, K.~Nordström, B.~Page, M.~Rüfenacht, M.~Schönherr and G.~Watt,
  %``LHAPDF6: parton density access in the LHC precision era,''
  Eur.\ Phys.\ J.\ C {\bf 75}, no. 3, 132 (2015)
  [arXiv:1412.7420 [hep-ph]].
  %%CITATION = ARXIV:1412.7420;%%
  %23 citations counted in INSPIRE as of 08 Oct 2015
\bibitem{Dobbs:2001ck}
  M.~Dobbs and J.~B.~Hansen,
  %``The HepMC C++ Monte Carlo event record for High Energy Physics,''
  Comput.\ Phys.\ Commun.\  {\bf 134}, 41 (2001).
  %%CITATION = CPHCB,134,41;%%
  %149 citations counted in INSPIRE as of 08 Oct 2015
\bibitem{Cullen:2014yla}
  G.~Cullen {\it et al.},
  %``G$\scriptsize{O}$S$\scriptsize{AM}$-2.0: a tool for automated one-loop calculations within the Standard Model and beyond,''
  Eur.\ Phys.\ J.\ C {\bf 74}, no. 8, 3001 (2014)
  [arXiv:1404.7096 [hep-ph]].
  %%CITATION = ARXIV:1404.7096;%%
  %36 citations counted in INSPIRE as of 08 Oct 2015
\bibitem{Cascioli:2011va}
  F.~Cascioli, P.~Maierhofer and S.~Pozzorini,
  %``Scattering Amplitudes with Open Loops,''
  Phys.\ Rev.\ Lett.\  {\bf 108}, 111601 (2012)
  [arXiv:1111.5206 [hep-ph]].
  %%CITATION = ARXIV:1111.5206;%%
  %175 citations counted in INSPIRE as of 08 Oct 2015
\bibitem{Kilian:2004pp} 
  W.~Kilian, D.~Rainwater and J.~Reuter,
  %``Pseudo-axions in little Higgs models,''
  Phys.\ Rev.\ D {\bf 71}, 015008 (2005)
  [hep-ph/0411213].
  %%CITATION = HEP-PH/0411213;%%
  %65 citations counted in INSPIRE as of 08 Oct 2015
\bibitem{Hagiwara:2005wg}
  K.~Hagiwara, W.~Kilian, F.~Krauss, T.~Ohl, T.~Plehn, D.~Rainwater, J.~Reuter and S.~Schumann,
  %``Supersymmetry simulations with off-shell effects for CERN LHC and ILC,''
  Phys.\ Rev.\ D {\bf 73}, 055005 (2006)
  [hep-ph/0512260].
  %%CITATION = HEP-PH/0512260;%%
  %71 citations counted in INSPIRE as of 08 Oct 2015
\bibitem{Beyer:2006hx}
  M.~Beyer, W.~Kilian, P.~Krstonosic, K.~Monig, J.~Reuter, E.~Schmidt and H.~Schroder,
  %``Determination of New Electroweak Parameters at the ILC - Sensitivity to New Physics,''
  Eur.\ Phys.\ J.\ C {\bf 48}, 353 (2006)
  [hep-ph/0604048].
  %%CITATION = HEP-PH/0604048;%%
  %44 citations counted in INSPIRE as of 08 Oct 2015
\bibitem{Alboteanu:2008my}
  A.~Alboteanu, W.~Kilian and J.~Reuter,
  %``Resonances and Unitarity in Weak Boson Scattering at the LHC,''
  JHEP {\bf 0811}, 010 (2008)
  [arXiv:0806.4145 [hep-ph]].
  %%CITATION = ARXIV:0806.4145;%%
  %66 citations counted in INSPIRE as of 08 Oct 2015
\bibitem{Kalinowski:2008fk}
  J.~Kalinowski, W.~Kilian, J.~Reuter, T.~Robens and K.~Rolbiecki,
  %``Pinning down the Invisible Sneutrino,''
  JHEP {\bf 0810}, 090 (2008)
  [arXiv:0809.3997 [hep-ph]].
  %%CITATION = ARXIV:0809.3997;%%
  %19 citations counted in INSPIRE as of 08 Oct 2015
\bibitem{Kilian:2014zja}
  W.~Kilian, T.~Ohl, J.~Reuter and M.~Sekulla,
  %``High-Energy Vector Boson Scattering after the Higgs Discovery,''
  Phys.\ Rev.\ D {\bf 91}, 096007 (2015)
  [arXiv:1408.6207 [hep-ph]].
  %%CITATION = ARXIV:1408.6207;%%
  %11 citations counted in INSPIRE as of 08 Oct 2015
\bibitem{Nason:2004rx}
  P.~Nason,
  %``A New method for combining NLO QCD with shower Monte Carlo algorithms,''
  JHEP {\bf 0411}, 040 (2004)
  [hep-ph/0409146].
  %%CITATION = HEP-PH/0409146;%%
  %976 citations counted in INSPIRE as of 08 Oct 2015
\bibitem{Frixione:2002ik}
  S.~Frixione and B.~R.~Webber,
  %``Matching NLO QCD computations and parton shower simulations,''
  JHEP {\bf 0206}, 029 (2002)
  [hep-ph/0204244].
  %%CITATION = HEP-PH/0204244;%%
  %1939 citations counted in INSPIRE as of 08 Oct 2015
\bibitem{Alwall:2014hca}
  J.~Alwall {\it et al.},
  %``The automated computation of tree-level and next-to-leading order differential cross sections, and their matching to parton shower simulations,''
  JHEP {\bf 1407}, 079 (2014)
  [arXiv:1405.0301 [hep-ph]].
  %%CITATION = ARXIV:1405.0301;%%
  %515 citations counted in INSPIRE as of 08 Oct 2015
\bibitem{Hoche:2010pf}
  S.~Hoche, F.~Krauss, M.~Schonherr and F.~Siegert,
  %``Automating the POWHEG method in Sherpa,''
  JHEP {\bf 1104}, 024 (2011)
  [arXiv:1008.5399 [hep-ph]].
  %%CITATION = ARXIV:1008.5399;%%
  %71 citations counted in INSPIRE as of 08 Oct 2015
\bibitem{Baer:2013cma}
  H.~Baer {\it et al.},
  %``The International Linear Collider Technical Design Report - Volume 2: Physics,''
  arXiv:1306.6352 [hep-ph].
  %%CITATION = ARXIV:1306.6352;%%
  %271 citations counted in INSPIRE as of 08 Oct 2015
\bibitem{Kilian:2006cj}
  W.~Kilian, J.~Reuter and T.~Robens,
  %``NLO Event Generation for Chargino Production at the ILC,''
  Eur.\ Phys.\ J.\ C {\bf 48}, 389 (2006)
  [hep-ph/0607127].
  %%CITATION = HEP-PH/0607127;%%
  %33 citations counted in INSPIRE as of 08 Oct 2015
\bibitem{Robens:2008sa}
  T.~Robens, J.~Kalinowski, K.~Rolbiecki, W.~Kilian and J.~Reuter,
  %``(N)LO Simulation of Chargino Production and Decay,''
  Acta Phys.\ Polon.\ B {\bf 39}, 1705 (2008)
  [arXiv:0803.4161 [hep-ph]].
  %%CITATION = ARXIV:0803.4161;%%
  %15 citations counted in INSPIRE as of 08 Oct 2015
\bibitem{Binoth:2009rv}
  T.~Binoth, N.~Greiner, A.~Guffanti, J.~Reuter, J.-P.~Guillet and T.~Reiter,
  %``Next-to-leading order QCD corrections to pp --> b anti-b b anti-b + X at the LHC: the quark induced case,''
  Phys.\ Lett.\ B {\bf 685}, 293 (2010)
  [arXiv:0910.4379 [hep-ph]].
  %%CITATION = ARXIV:0910.4379;%%
  %84 citations counted in INSPIRE as of 09 Oct 2015
\bibitem{Greiner:2011mp}
  N.~Greiner, A.~Guffanti, T.~Reiter and J.~Reuter,
  %``NLO QCD corrections to the production of two bottom-antibottom pairs at the LHC,''
  Phys.\ Rev.\ Lett.\  {\bf 107}, 102002 (2011)
  [arXiv:1105.3624 [hep-ph]].
  %%CITATION = ARXIV:1105.3624;%%
  %63 citations counted in INSPIRE as of 09 Oct 2015
\bibitem{Kinoshita:1962ur}
  T.~Kinoshita,
  %``Mass singularities of Feynman amplitudes,''
  J.\ Math.\ Phys.\  {\bf 3}, 650 (1962).
  %%CITATION = JMAPA,3,650;%%
  %1173 citations counted in INSPIRE as of 08 Oct 2015
\bibitem{Lee:1964is}
  T.~D.~Lee and M.~Nauenberg,
  %``Degenerate Systems and Mass Singularities,''
  Phys.\ Rev.\  {\bf 133}, B1549 (1964).
  %%CITATION = PHRVA,133,B1549;%%
  %1026 citations counted in INSPIRE as of 08 Oct 2015
\bibitem{Catani:1996vz}
  S.~Catani and M.~H.~Seymour,
  %``A General algorithm for calculating jet cross-sections in NLO QCD,''
  Nucl.\ Phys.\ B {\bf 485}, 291 (1997)
  [Nucl.\ Phys.\ B {\bf 510}, 503 (1998)]
  [hep-ph/9605323].
  %%CITATION = HEP-PH/9605323;%%
  %1241 citations counted in INSPIRE as of 08 Oct 2015
\bibitem{Frixione:1995ms}
  S.~Frixione, Z.~Kunszt and A.~Signer,
  %``Three jet cross-sections to next-to-leading order,''
  Nucl.\ Phys.\ B {\bf 467}, 399 (1996)
  [hep-ph/9512328].
  %%CITATION = HEP-PH/9512328;%%
  %517 citations counted in INSPIRE as of 08 Oct 2015
\bibitem{Frixione:2009yq}
  R.~Frederix, S.~Frixione, F.~Maltoni and T.~Stelzer,
  %``Automation of next-to-leading order computations in QCD: The FKS subtraction,''
  JHEP {\bf 0910}, 003 (2009)
  [arXiv:0908.4272 [hep-ph]].
  %%CITATION = ARXIV:0908.4272;%%
  %155 citations counted in INSPIRE as of 08 Oct 2015
\bibitem{Bevilacqua:2011xh}
  G.~Bevilacqua, M.~Czakon, M.~V.~Garzelli, A.~van Hameren, A.~Kardos, C.~G.~Papadopoulos, R.~Pittau and M.~Worek,
  %``Helac-nlo,''
  Comput.\ Phys.\ Commun.\  {\bf 184}, 986 (2013)
  [arXiv:1110.1499 [hep-ph]].
  %%CITATION = ARXIV:1110.1499;%%
  %158 citations counted in INSPIRE as of 08 Oct 2015
\bibitem{Alioli:2010xd}
  S.~Alioli, P.~Nason, C.~Oleari and E.~Re,
  %``A general framework for implementing NLO calculations in shower Monte Carlo programs: the POWHEG BOX,''
  JHEP {\bf 1006}, 043 (2010)
  [arXiv:1002.2581 [hep-ph]].
  %%CITATION = ARXIV:1002.2581;%%
  %834 citations counted in INSPIRE as of 08 Oct 2015
\bibitem{Muck:2015cra}
  A.~M\"uck,
  %``Parton-shower matching for electroweak corrections,''
  Nucl.\ Part.\ Phys.\ Proc.\  {\bf 261-262}, 308.
\bibitem{Binoth:2010xt}
  T.~Binoth {\it et al.},
  %``A Proposal for a standard interface between Monte Carlo tools and one-loop programs,''
  Comput.\ Phys.\ Commun.\  {\bf 181}, 1612 (2010)
  [arXiv:1001.1307 [hep-ph]].
  %%CITATION = ARXIV:1001.1307;%%
  %87 citations counted in INSPIRE as of 08 Oct 2015
\bibitem{Alioli:2013nda}
  S.~Alioli {\it et al.},
  %``Update of the Binoth Les Houches Accord for a standard interface between Monte Carlo tools and one-loop programs,''
  Comput.\ Phys.\ Commun.\  {\bf 185}, 560 (2014)
  [arXiv:1308.3462 [hep-ph]].
  %%CITATION = ARXIV:1308.3462;%%
  %33 citations counted in INSPIRE as of 08 Oct 2015
\bibitem{Nogueira:1991ex}
  P.~Nogueira,
  %``Automatic Feynman graph generation,''
  J.\ Comput.\ Phys.\  {\bf 105}, 279 (1993).
  %%CITATION = JCTPA,105,279;%%
  %557 citations counted in INSPIRE as of 08 Oct 2015
\bibitem{Kuipers:2012rf}
  J.~Kuipers, T.~Ueda, J.~A.~M.~Vermaseren and J.~Vollinga,
  %``FORM version 4.0,''
  Comput.\ Phys.\ Commun.\  {\bf 184}, 1453 (2013)
  [arXiv:1203.6543 [cs.SC]].
  %%CITATION = ARXIV:1203.6543;%%
  %93 citations counted in INSPIRE as of 08 Oct 2015
\bibitem{Ossola:2006us}
  G.~Ossola, C.~G.~Papadopoulos and R.~Pittau,
  %``Reducing full one-loop amplitudes to scalar integrals at the integrand level,''
  Nucl.\ Phys.\ B {\bf 763}, 147 (2007)
  [hep-ph/0609007].
  %%CITATION = HEP-PH/0609007;%%
  %437 citations counted in INSPIRE as of 08 Oct 2015
\bibitem{Mastrolia:2008jb}
  P.~Mastrolia, G.~Ossola, C.~G.~Papadopoulos and R.~Pittau,
  %``Optimizing the Reduction of One-Loop Amplitudes,''
  JHEP {\bf 0806}, 030 (2008)
  [arXiv:0803.3964 [hep-ph]].
  %%CITATION = ARXIV:0803.3964;%%
  %101 citations counted in INSPIRE as of 08 Oct 2015
\bibitem{Denner:2012yc}
  A.~Denner, S.~Dittmaier, S.~Kallweit and S.~Pozzorini,
  %``NLO QCD corrections to off-shell top-antitop production with leptonic decays at hadron colliders,''
  JHEP {\bf 1210}, 110 (2012)
  [arXiv:1207.5018 [hep-ph]].
  %%CITATION = ARXIV:1207.5018;%%
  %55 citations counted in INSPIRE as of 08 Oct 2015
\bibitem{Lei:2008ii}
  L.~Guo, W.~G.~Ma, R.~Y.~Zhang and S.~M.~Wang,
  %``One-loop QCD corrections to the e+ e- ---> W+ W- b anti-b process at the ILC,''
  Phys.\ Lett.\ B {\bf 662}, 150 (2008)
  [arXiv:0802.4124 [hep-ph]].
  %%CITATION = ARXIV:0802.4124;%%
  %3 citations counted in INSPIRE as of 08 Oct 2015
\bibitem{Jezabek:1988iv}
  M.~Jezabek and J.~H.~Kuhn,
  %``QCD Corrections to Semileptonic Decays of Heavy Quarks,''
  Nucl.\ Phys.\ B {\bf 314}, 1 (1989).
  %%CITATION = NUPHA,B314,1;%%
  %309 citations counted in INSPIRE as of 08 Oct 2015
\bibitem{Hoang:2011it}
  M.~Stahlhofen and A.~Hoang,
  %``NNLL top-antitop production at threshold,''
  PoS RADCOR {\bf 2011}, 025 (2011)
  [arXiv:1111.4486 [hep-ph]].
  %%CITATION = ARXIV:1111.4486;%%
  %10 citations counted in INSPIRE as of 08 Oct 2015
\bibitem{Bach:2014hca}
  F.~Bach and M.~Stahlhofen,
  %``Top pair threshold production at a linear collider with WHIZARD,''
  arXiv:1411.7318 [hep-ph].
  %%CITATION = ARXIV:1411.7318;%%
  %1 citations counted in INSPIRE as of 08 Oct 2015
\bibitem{Cacciari:2008gp}
  M.~Cacciari, G.~P.~Salam and G.~Soyez,
  %``The Anti-k(t) jet clustering algorithm,''
  JHEP {\bf 0804}, 063 (2008)
  [arXiv:0802.1189 [hep-ph]].
  %%CITATION = ARXIV:0802.1189;%%
  %3080 citations counted in INSPIRE as of 08 Oct 2015
\bibitem{Buckley:2010ar}
  A.~Buckley, J.~Butterworth, L.~Lonnblad, D.~Grellscheid, H.~Hoeth, J.~Monk, H.~Schulz and F.~Siegert,
  %``Rivet user manual,''
  Comput.\ Phys.\ Commun.\  {\bf 184}, 2803 (2013)
  [arXiv:1003.0694 [hep-ph]].
  %%CITATION = ARXIV:1003.0694;%%
  %195 citations counted in INSPIRE as of 08 Oct 2015
\bibitem{Chokoufe:2015xx} B. Chokouf\'e Nejad, W. Kilian, J.R. Reuter, C. Weiss:
	PoS(EPS-HEP2015)317, (2015).

\end{thebibliography}
\end{document}